\documentclass[11pt]{article}
\usepackage{amssymb,amsthm,amsmath,color,graphics}
\theoremstyle{plain}
\newtheorem{theorem}{Theorem}

\newtheorem{remark}[theorem]{Remark}
\newtheorem{conjecture}[theorem]{Conjecture}
\theoremstyle{definition}
\newtheorem{definition}{Definition}

\newcommand{\be}{\begin{equation}}
\newcommand{\ee}{\end{equation}}
\newcommand{\bea}{\begin{eqnarray}}
\newcommand{\eea}{\end{eqnarray}}
\newcommand{\bc}{\begin{center}}
\newcommand{\ec}{\end{center}}

\newcommand{\OOB}{\textsf{OOB}}

\newcommand{\PGE}{\textsf{PGE}}

\newcommand{\SEB}{\textsf{wc-AKE}}

\newcommand{\SKE}{\textsf{sc-AKE}}
\newcommand{\scAKE}{\textsf{sc-AKE}}

\newcommand{\qAKE}{\textsf{q-AKE}}

\newcommand{\SIGQKE}{\qAKE$_{\textsf{pub}}$}
\newcommand{\MACQKE}{\qAKE$_{\textsf{sym}}$}

\newcommand{\SIGSKE}{\scAKE$_{\textsf{pub}}$}
\newcommand{\MACSKE}{\scAKE$_{\textsf{sym}}$}

\newcommand{\cUKE}{\textsf{c-UKE}}
\newcommand{\qUKE}{\textsf{q-UKE}}

\def\opone{\leavevmode\hbox{\small1\kern-3.8pt\normalsize1}}

\begin{document}

\medskip

\bc

{\Large \bf A new spin on quantum cryptography:\\ Avoiding trapdoors
and embracing public keys}

\medskip

\renewcommand{\thefootnote}{\fnsymbol{footnote}}
{\footnotesize Lawrence M. Ioannou$^{1,2}$ and Michele
Mosca$^{1,2,3}$ \\ {\it
$^1$Institute for Quantum Computing, University of Waterloo,\\
200 University Avenue, Waterloo, Ontario, N2L 3G1, Canada \\
$^2$Department of Combinatorics and Optimization, University of Waterloo,\\
200 University Avenue, Waterloo, Ontario, N2L 3G1, Canada\\
$^3$Perimeter Institute for Theoretical Physics\\31 Caroline Street
North, Waterloo, Ontario, N2L 2Y5, Canada} }

\ec

\begin{quote}{\small
We give new arguments in support of \emph{signed quantum key
establishment}, where quantum cryptography is used in a public-key
infrastructure that provides the required authentication.  We also
analyze more thoroughly than previous works the benefits that
quantum key establishment protocols have over certain classical
protocols, motivated in part by the various objections to quantum
key establishment that are sometimes raised.  Previous knowledge of
quantum cryptography on the reader's part is not required for this
article, as the definition of ``quantum key establishment'' that we
use is an entirely classical and black-box characterization (one
need only trust that protocols satisfying the definition exist).
}
\end{quote}

Quantum cryptography\footnote{Note that quantum cryptography
includes many protocols that this paper does not discuss.  We use
the term ``quantum cryptography'' here as a synonym for ``quantum
key establishment'', often called ``quantum key distribution'' or
``\textsc{qkd}''.} has been promoted as a more secure alternative to
public-key cryptography based on computational assumptions (see the
abstract of Ref. \cite{BS99} for a typical example). However, an
opposing view is sometimes voiced by classical cryptographers and
computer security specialists questioning whether quantum
cryptography is really a practical way to achieve security against
quantum computers, also known as \emph{quantum resistance}. Several
detailed analyses have appeared that consider the benefits and
disadvantages of quantum cryptography in comparison to classical
alternatives \cite{PPS07,SECOQC:White:09,SML09,Ber09}. The present
article contributes to the dialogue in a way that we hope is very
palatable to the community of quantum-questioning cryptographers: we
give new arguments in support of \emph{signed quantum key
establishment}, where quantum cryptography is used in a public-key
infrastructure that provides the required authentication.

We also analyze more thoroughly than previous works the benefits
that quantum key establishment (\textsc{qke}) protocols have over
certain classical protocols, motivated in part by the various
objections to \textsc{qke} that have been put forward (for example,
in Ref. \cite{Ber09}). Some of those objections follow.\footnote{We
have stated these objections in our own words.}

\begin{itemize}\label{objections}
\item \textbf{Objection 1}: Quantum computers are not known to be able to break all
classical public-key cryptosystems, such as the McEliece
cryptosystem or those based on lattice problems; so we can just
upgrade to these quantum-resistant cryptosystems and forget quantum
cryptography---that way, we'd retain all the benefits of a
public-key infrastructure.

\item \textbf{Objection 2}: If all of classical public-key cryptography is found to be
easily breakable, then we might as well revert to using our best
symmetric-key cryptography, including block ciphers like
\textsc{aes}, which we all agree is quantum resistant; quantum
cryptography would require symmetric shared initial keys anyway in
this case, so it wouldn't gain us anything.

\item \textbf{Objection 3}: We don't need any means of key distribution,
let alone a quantum mechanical one---let's just exchange a
lifetime's worth of symmetric keying material at the start.  If for
whatever reason we do need new keys, see Objection 4.

\item \textbf{Objection 4}: We don't need any means of generating independent
secret key over telecommunication links---let's just use a trusted
courier each time we need independent secret key.
\end{itemize}

\noindent We address all of these objections.

\paragraph{\textbf{{Not quantum cryptography again.}}}

Like in pro-quantum-cryptography articles that have come before
this, we assume here that the universe is quantum mechanical, so
that, at a minimum, the secret key generated by a secure
key-establishment protocol must be secure against an adversary able
to perform probabilistic-polynomial-time computations on a quantum
computer. As well, as stated by Stebila et al. \cite{SML09}, we
``expect the costs and challenges of using [\textsc{qke}] to
decrease to the point where [such] systems can be deployed
affordably and their behaviour can be certified.'' In fact, most of
the advantages of quantum cryptography that we point out here have
been noted by Paterson et al. \cite{PPS07} or Stebila et al.
\cite{SML09}.

Despite these similarities to previous works, our analysis contains
distinct new features: it

\begin{itemize}

\item suggests a new way to define the classes of classical and \textsc{qke}
protocols, in order to aid their comparison,

\item deals properly with the option of using trusted couriers instead of \textsc{qke},
by distinguishing between in-band and out-of-band actions,

\item uses the weakest possible notion of ``security'' in a quantum universe (i.e. computational security), and therefore does not
focus on information-theoretic security---for its own sake---as an
advantage of \textsc{qke} over computationally-secure classical
alternatives,

\item provides a finer-grained analysis of the computational
assumptions underlying the classical alternatives to \textsc{qke},

\item highlights a property (we call it ``nonattributability'') of
\textsc{qke} that has received little attention in the literature,
and

\item supports a recommendation that is both theoretically and
practically sound, which both sides of the ``quantum debate'' can
agree upon.
\end{itemize}

\noindent Generally, we hope the reader finds this article to
benefit from a more precise cryptographic analysis, despite its more
limited scope in taking an idealized view and thus not discussing
the more technological or economical aspects of \textsc{qke}
(including side-channel attacks). In other words, this paper studies
the value of the \textsc{qke} primitive assuming it is available in
practice and is as cost-effective as any type of ``in-band''
classical key establishment (see Definition
\ref{def_OutOfBand}).\footnote{ The practical availability of the
\textsc{qke} primitive between a typical real-world Alice and Bob is
a very non-trivial assumption. For a fairly recent status report on
practical \textsc{qke} systems, one can see Ref. \cite{LS09}, where
it is evident that key-rate, distance and availability remain
serious obstacles for most practical applications today. In the
cases that one believes that \textsc{qke} could in principle add
value, one will need to do an in depth analysis of the various costs
and practical limitations before deciding whether in some particular
practical situation \textsc{qke} will be the preferred alternative.
Weighing the costs against the value depends on many parameters
which vary widely from place to place and over time, and analyzing
this broad spectrum is beyond the scope of this paper.}~~We adopt
the same foundational approach that Goldreich does in Refs.
\cite{Gol01,Gol04}. This basically means that, when reviewing which
computational assumptions are known to be necessary or sufficient
for certain cryptographic primitives, we ignore those assumptions
(and the schemes based on them) that are ad hoc: we deal only in
fundamental computational assumptions, in particular, one-way
functions and trapdoor predicates.

But the foregoing analysis is not as complete as it could be. In
particular, we do not treat the distributed authenticated key
establishment problem (i.e., in a network setting and where
simultaneous, multiple key establishment sessions among many pairs
of users are considered) as rigorously as it deserves (e.g.
\cite{BR94,CK01}). That is, we implicitly assume that
\emph{point-to-point}\footnote{By ``point-to-point'' protocols or
key establishment systems we mean those that presume a unique pair
of honest participants in the protocol; in other words, Alice and
Bob are fixed.} unauthenticated key establishment protocols (whether
they be key transport protocols or key agreement
protocols\footnote{Recall that a \emph{key transport protocol} is a
key establishment protocol where the final secret key is generated
by one party and sent to the other party (using some kind of
encryption mechanism). By contrast, a \emph{key agreement protocol}
is a key establishment protocol where both parties contribute to the
generation of the final secret key. See Ref. \cite{MvOV96} for more
details.}) and message-authentication protocols (whether they be
digital signature schemes or message authentication codes) may be
combined in such a way as to form robust \emph{distributed}
authenticated key establishment protocols, without stating the
details of how this combining---especially with regard to
authentication---actually works.\footnote{We follow Ref.
\cite{MvOV96} in our use of the terms ``authenticated (key
establishment)'' and ``unauthenticated (key establishment)''.  In
this convention, the word ``(un)authenticated'' describes the
\emph{guaranteed condition} of the final shared key resulting from
the protocol. We note that this convention is the opposite of that
in Ref. \cite{Gol04}, where ``(un)authenticated'' describes the
\emph{a priori assumption} on the (classical) communication channel
used in the protocol.}~ This
 deficiency is manifest in the definition of ``security'' that
we use (Definition \ref{def_Q-Resistant}): it only refers to privacy
of the secret key and not its integrity; we take authentication
\emph{in a network-setting} for granted (for both classical and
quantum networks). Thus, analyzing point-to-point key establishment
systems is sufficient for our scope and, for such systems, integrity
of the established secret key is obtained either by assumption (in
the case of unauthenticated key establishment) or by the
message-authentication protocols used to authenticate the classical
communication channel (in the case of authenticated key
establishment). Our omission of the analysis of distributed
\textsc{qke} in no way is meant to imply that the problem is
trivial---we believe it is an important open problem, which to our
knowledge has not been addressed in any previous works.

As a final note to the reader, we stress that previous knowledge of
quantum cryptography is not required for this article.  The
definition of ``\textsc{qke}'' that we use is an entirely classical
and black-box characterization (one need only trust that protocols
satisfying the definition exist).

\paragraph{\textbf{{Key establishment.}}}
We are ultimately interested in authenticated key establishment (or
\textsc{ake}), since, in practice, it is usually not a reasonable
assumption that the classical channel connecting Alice and Bob is
authenticated a priori.  But we shall also consider unauthenticated
key establishment (or \textsc{uke}), because, as well as being
useful as a building block for \textsc{ake} systems, it is an
often-considered cryptographic primitive in more foundational works,
e.g., Ref. \cite{IR88} (see Remark \ref{rem_SKA}). We now make some
precise definitions.

A (point-to-point) \emph{\textsc{ake} system} consists of two
probabilistic-polynomial-time (quantum) computers, called ``Alice''
and ``Bob'', that
\begin{itemize}

\item are preloaded with classical \emph{initial keys}, $k_A$ (stored on Alice) and $k_B$ (stored on Bob), which
are pre-distributed out of band (see Definition \ref{def_OutOfBand})
in an authenticated and, where necessary (for example, when the keys
are symmetric), private fashion, and

\item are connected by two insecure channels, one quantum and one classical,
variously monitored or controlled by an adversarial
probabilistic-polynomial-time (quantum) computer, called ``Eve'',
and

\item together execute a particular (point-to-point) \emph{\textsc{ake} protocol},
the specification $\pi$ of which is preloaded authentically but is
not secret, and

\item which results in Alice and Bob computing
outputs $s_A$ and $s_B$, respectively, such that either $s_A = s_B
 = \bot$, which corresponds to Alice and Bob aborting the protocol, or
$s_A$ and $s_B$ are bit-strings, in which case, if $s_A = s_B$, then
the \emph{secret key} $s := s_A$ is defined.

\end{itemize}

\noindent 
When the initial keys are symmetric ($k_A = k_B$), we may use $k$ to
denote each one, i.e., $k = k_A = k_B$; if the initial keys are
asymmetric ($k_A \neq k_B$), then
\begin{eqnarray}
k_A &=& (x_A,y_B)\\
k_B &=& (x_B, y_A),
\end{eqnarray}
where $(x_A, y_A)$ is Alice's private-public key-pair and $(x_B,
y_B)$ is Bob's private-public key-pair. We will say more about
asymmetric (public-key) cryptography later on.

\begin{definition}[In band/out of band]\label{def_OutOfBand}  The term ``in
band'' describes actions carried out in the normal course of
telecommunications strictly via remote signalling across
communication channels.  The term ``out of band'' is used to mean
``not in band'' and describes communication via non-digital/manual
means as opposed to via standard telecommunication devices.
\end{definition}

\begin{remark}[Classical channel]  Strictly speaking, there is
no need for a dedicated classical channel between Alice and Bob,
since classical information can be sent along the quantum channel.
However, the well-known \textsc{qke} protocols (i.e., those based on
the ones in Refs \cite{BB84,Eke91}) clearly distinguish the
classical from the quantum communication; in particular, it suffices
that only the classical communication is authenticated in order for
the secret key to be authenticated at the end of the protocol
(whereas, one could imagine a quantum protocol where the quantum
communication also needs to be authenticated).  In line with this
distinction, we assume separate quantum and classical channels.
\end{remark}

A (point-to-point) \emph{\textsc{uke} system} is defined similarly
to an \textsc{ake} system, with only the following differences:
\begin{itemize}
\item Alice and Bob possess no initial keys and
\item the classical channel is assumed to be authenticated, i.e., Eve
is assumed only to passively monitor the classical channel (but she
can still totally control the quantum channel), and
\item $\pi$ is a (point-to-point)
\emph{\textsc{uke} protocol}.
\end{itemize}

We also need to define conditions under which a key establishment
protocol is secure or, more specifically, quantum-resistant. We
would like a definition that applies equally well to both quantum
and fully classical protocols, i.e., all protocols allowed in the
above frameworks.  Since we take authentication for granted (as
explained above), the following security definition is sufficient
for both \textsc{ake} and \textsc{uke} systems.  Call a key
establishment protocol \emph{perfectly secure} if, for any algorithm
for Eve, we have that (1) $s_A = s_B$, (2) if $s_A \neq \bot$ then
$s_A$ is uniformly distributed and independent of Eve's state, and
(3) if Eve does not interfere with the protocol (where we assume
otherwise perfect channels), then $s_A \neq \bot$.  Let
$\mathcal{I}$ be an \emph{ideal key establishment system} that
implements a perfectly secure protocol.  Let $\mathcal{R}(\pi)$ be a
\emph{real key establishment system} that uses protocol $\pi$.  Let
$n$ be the minimum length of the secret key $s$ if Alice and Bob do
not abort.  Consider a probabilistic-polynomial-time (quantum)
 \emph{distinguisher} running in time polynomial in $n$, that
interacts with either $\mathcal{I}$ or $\mathcal{R}(\pi)$ and then
outputs a guess bit $B$; the distinguisher has access to Eve's
system and the outputs $s_A$ and $s_B$.

\begin{definition}[Quantum-resistant key-establishment protocol (with respect to privacy)]\label{def_Q-Resistant}
Assuming the above definitions, a point-to-point key-establishment
protocol $\pi$ is \emph{quantum-resistant (with respect to privacy)}
if, for any such distinguisher, the quantity
\begin{eqnarray}
|\Pr[B=1 | \mathcal{I}] - \Pr[B=1|\mathcal{R}(\pi)]|
\end{eqnarray}
is negligible for all sufficiently large $n$, where $\Pr[B=1 |
\mathcal{I}]$ and $\Pr[B=1|\mathcal{R}(\pi)]$ are the probabilities
that $B=1$ when the distinguisher interacts with $\mathcal{I}$ and
$\mathcal{R}(\pi)$, respectively.
\end{definition}
\noindent We give this (semi-formal) definition for completeness; we
refer the reader to Refs \cite{MQR09,Can00,Gol01,NC00} for how to
rigorize such a definition.

As a final specification of our basic setup, it will be helpful to
define the \emph{classical communication} $c$ in a key establishment
protocol. For classical protocols, the classical communication is
all the communication between Alice and Bob.  For arbitrary
(quantum) protocols, defining the classical communication is a bit
more subtle; we refrain from giving a formal definition here (for
the sake of the reader who may be unfamiliar with quantum
measurement). Rather, for the quantum protocols we care about, it
suffices to define the classical communication tautologically as the
classical communication specified in the protocol, since these
protocols clearly and naturally distinguish the classical and
quantum information sent between Alice and Bob.

\paragraph{\textbf{{The contenders.}}}
Below are listed and defined two main classes of point-to-point
\textsc{uke} protocols as well as the five main classes of
point-to-point \textsc{ake} protocols that are considered in the
literature when evaluating the usefulness of quantum cryptography in
comparison to classical techniques for key establishment.  These
classes, as defined, do not cover all conceivable protocols, but do
cover all the ones that are usually considered (which suffices
here). In defining these classes, we restrict to quantum-resistant
protocols (because the universe is quantum). It will help to view
the quantities $k_A$, $k_B$, $k$, $s$, and $c$ introduced above as
random variables.  For example, in the case of symmetric initial
keys, the quantity $k$ may be viewed as a uniformly distributed
random variable in $\{0,1\}^\ell$, for some fixed $\ell \in
\mathbb{Z}^{>0}$ that determines the length of the initial keys.
\\
 \\
\noindent\emph{Unauthenticated key establishment protocols:}

\begin{itemize}
\item \emph{Classical \textsc{uke}} (\cUKE)---This class includes any
quantum-resistant and totally classical \textsc{uke} protocol.
  It includes unauthenticated key transport protocols based on
public-key encryption (but not those based on symmetric-key
encryption).

\item \emph{Quantum \textsc{uke}} (\qUKE)---This
class includes any quantum-resistant \textsc{uke} protocol such
that, whenever Eve has not interfered with the protocol,
the secret key $s$ 
is independent of the
classical communication $c$, i.e., 
for all values $c'$ of the classical communication and all values
$s'$ of the secret key,
\begin{eqnarray}\label{eqn_qUKE}
\Pr[s=s'|c=c']&=&\Pr[s=s'].
\end{eqnarray}
It includes the well-known \textsc{qke} protocols and can easily be
shown not to include any classical protocols.\footnote{We sketch a
proof of the latter fact that no purely classical protocol can be
quantum resistant and satisfy (\ref{eqn_qUKE}). Let $r_A$ and $r_B$
be binary strings encoding the private local randomness that Alice
and Bob respectively use in the protocol. Consider the sequence
$c_1, c_2, \ldots$ of messages passed between Alice and Bob.  Each
$c_i$ places constraints on the values of $r_A$ and $r_B$.  Since,
at the end of the protocol, the secret key $s$ is uniquely
determined, it must be that $r_A$ and $r_B$ are determined by the
classical communication $c$ up to implying a unique $s$, i.e.,
$H(s|c)=0$, where $H$ is the Shannon entropy. For any two random
variables $X$ and $Y$, $H(X|Y) = H(X)$ if and only if $X$ and $Y$
are independent \cite{Sti95}. Therefore, if (\ref{eqn_qUKE}) holds,
then $H(s)=H(s|c)=0$, so that $s$ is a constant and thus the
protocol is not quantum resistant.}
\end{itemize}

\begin{remark}[Secret key agreement]\label{rem_SKA} The cryptographic primitive
realized by protocols in \emph{\cUKE} is usually referred to as
\emph{secret key agreement} (or sometimes just \emph{secret
agreement}) in the literature.  Note that this primitive is also
realized by protocols in \emph{\qUKE}.
\end{remark}

\noindent\emph{Authenticated key establishment protocols:}

\begin{itemize}

\item \emph{Out-of-band key establishment} (\OOB)---This class includes any \textsc{ake} protocol where Alice and
Bob are preloaded with the secret key out of band, i.e.,
\begin{eqnarray}
s = k_A = k_B.
\end{eqnarray}
\noindent It includes protocols that employ a trusted courier.  The
initial keys in such protocols are typically much larger than in
protocols belonging to the classes below.

\item \emph{Pseudorandom generator expansion} (\PGE)---This
class includes any\\ quantum-resistant and totally classical
\textsc{ake} protocol not in \OOB~that uses symmetric initial keys
where Alice and Bob establish a secret key that
is efficiently computable from the initial keys, i.e.,
there exists a deterministic-polynomial-time classical algorithm $A$
such that
\begin{eqnarray}
s = A(\pi, k).
\end{eqnarray}
It includes protocols that use a pseudorandom generator to expand
the initial keys into a secret key.

\item \emph{Weak classical \textsc{ake}}
(\SEB)---This class includes any quantum-resistant and totally
classical \textsc{ake} protocol in neither \PGE~nor \OOB~that uses
symmetric initial keys.   Note such protocols have the property that
the secret key
is efficiently computable from the initial keys and the
communication, i.e., 
there exists a deterministic-polynomial-time classical algorithm $A$
such that
\begin{eqnarray}
s = A(\pi, k , c).
\end{eqnarray}
  The class includes authenticated key transport protocols based on
symmetric-key encryption.

\item \emph{Strong\footnote{Our use of the word ``strong'' differs from
that in Ref. \cite{LLM07}, where a key establishment protocol is
secure only if it remains secure under the reveal of any subset of
the initial (also called ``long-term'') and ephemeral keys that does
not contain both the initial and ephemeral keys of one of the
parties. The protocols of the class we define here need only remain
secure under the reveal of the initial keys.  Indeed, the ``strong''
of Ref. \cite{LLM07} is stronger than ours.} classical \textsc{ake}}
(\scAKE)---This class includes any quantum-resistant and totally
classical \textsc{ake} protocol, where Alice and Bob establish an
authenticated secret key $s$ that is not functionally dependent on
the initial keys $k_A$ and $k_B$, i.e., there exists a
deterministic-polynomial-time classical algorithm $A$ such that
\begin{eqnarray}\label{eqn_scAKE}
s = A(\pi, r_A , r_B),
\end{eqnarray}
where $r_A$ and $r_B$ are (random variables representing) the
private local random choices of Alice and Bob respectively (made
independently of the initial keys). It includes authenticated key
transport protocols based on public-key encryption (but not those
based on symmetric-key encryption); more generally, it includes the
``authenticated version'' of any quantum-resistant \textsc{uke}
protocol, where the initial keys are used (only) to authenticate all
the communication of the protocol (see Remark
\ref{rem_UKEimpliesAKE}).

\item \emph{Quantum \textsc{ake}} (\qAKE)---This
class includes any quantum-resistant \textsc{ake} protocol such
that, whenever Eve
has not interfered with the protocol, the secret key $s$ 
is independent of the initial keys and the
classical communication $c$, i.e., 
for all values $k_A'$ and $k_B'$ of the initial keys and all values
$c'$ of the classical communication and all values $s'$ of the
secret key,
\begin{eqnarray}\label{eqn_qAKE}
\Pr[s=s'|k_A = k_A' , k_B = k_B', c=c']&=&\Pr[s=s'].
\end{eqnarray}
It includes the authenticated version of the well-known \textsc{qke}
protocols and can easily be shown not to include any classical
protocols (similarly to the class \qUKE, defined previously).
\end{itemize}

\begin{remark}[Possible emptiness of classical classes]
Of the classes of in-band key establishment protocols, only
{\em{\qUKE}} and \em{\qAKE} are known to be nonempty.
\end{remark}

\begin{remark}[Key pre-distribution v. dynamic key establishment]  The union of the classes \emph{\OOB}~and
\emph{\PGE}~contains protocols referred to collectively as \emph{key
pre-distribution schemes} \cite{MvOV96}, which is why we label these
two classes differently.  Note that there is no need to authenticate
the in-band communication in these protocols because there is none.
Protocols that are not key pre-distribution schemes are said to
accomplish \emph{dynamic key establishment}.
\end{remark}

\begin{remark}[Definition of \SKE]
The class \emph{\SKE}~may contain protocols that use the ``quantum
public-key cryptosystems'' in Ref. \cite{OTU00}, since the model
does not stipulate how initial keys are derived (i.e., they could be
derived using a quantum computer). 
\end{remark}

\begin{remark}[Definition of \qAKE]
The class \emph{\qAKE}~ may contain protocols obeying physical
theories other than quantum theory.
\end{remark}

\begin{remark}[\textsc{uke} implies \textsc{ake}]\label{rem_UKEimpliesAKE} Note that if $\pi$ is in \emph{\cUKE}, then $\pi$ naturally gives
rise to a protocol in \emph{\scAKE}~ when combined with a secure
classical message-authentication protocol.  A similar statement
holds for \emph{\qUKE} and \emph{\qAKE}.
\end{remark}

We subdivide the classes \SKE~ and \qAKE~ by the type of initial
keys---either symmetric or public---used in the particular key
establishment protocol, i.e., we have the following disjoint unions
\begin{eqnarray}
\textrm{\SKE} &=& \textrm{\MACSKE} \cup \textrm{\SIGSKE} \\
\textrm{\qAKE} &=& \textrm{\MACQKE} \cup \textrm{\SIGQKE}.
\end{eqnarray}

\noindent Table 1 summarizes the different classes by the various
categories.

\begin{table}[hbt]\label{tbl_CompAssumptions} \centering
\begin{tabular}{|c||c|c||c|}
\hline
 & \textsc{uke} & \textsc{ake} & \\
\hline \hline key pre-distribution & - & \OOB & out-of-band\\
\cline{2-4}
 & - & \PGE & in-band\\
\cline{1-3}
dynamic & - & \SEB & \\
key establishment & \cUKE  & \scAKE &\\
& \qUKE  & \qAKE &\\
\hline
\end{tabular}
\caption{The different classes of key establishment protocols.}
\end{table}

\paragraph{\textbf{{Apples and Oranges.}}}
The class \OOB~ is included in the above list (and in the following
analysis) largely for completeness; it is not technically considered
a key establishment protocol. Out-of-band protocols for key
establishment need not employ any fundamental cryptographic
primitives and cannot provide the same essential functionality that
in-band protocols do, i.e., generating new secret key in band. The
generally accepted view is that out-of-band key establishment is the
most secure way to establish potentially very long secret keys, but
that well-implemented in-band protocols typically provide either a
more feasible solution in particular applications or a more
cost-effective solution in the long term.  Because we are making the
(reasonable) assumption that \textsc{qke} will be cost-effective in
the future, it reasonably follows that, in at least some cases, it
will also be more cost-effective than out-of-band key establishment
in the future. We mean to challenge here previous comments made by
Bernstein \cite{Ber09}, that trusted couriers perform equally as
well as \textsc{qke} systems insofar as their ability to generate
entropy in the cryptographic system (from Eve's point of view). The
distinction between in-band and out-of-band entropy generation is an
important one (cost-wise), and it is impossible to generate entropy
in band using classical cryptography alone.

\paragraph{\textbf{{Computational assumptions.}}}
We would like to closely examine the fundamental computational
assumptions that underlie the various kinds of key establishment
protocols.  To do this, we start by recalling the following
well-known theorems.\footnote{The following theorems and other
similar statements should be interpreted as follows.  A statement of
the form ``Cryptographic objects of type $Y$ exist if cryptographic
objects of type $X$ exist'' means ``If there exists an object of
type $X$, then there exists an object of type $Y$ such that breaking
the object of type $Y$ implies breaking the object of type $X$.''
Such a statement may also be phrased, ``$X$ implies $Y$''.}

\begin{theorem}[\cite{Gol01}]\label{thm_PGiffOWF}  Pseudorandom generators exist if and only if one-way functions exist.
\end{theorem}

\begin{theorem}[\cite{Gol04}]\label{thm_SKEiffOWF}  Symmetric-key encryption schemes exist if and only if
 one-way functions exist.
\end{theorem}


\begin{theorem}[\cite{GM84}]\label{thm_PKEifTP}  Public-key encryption schemes exist
if and only if
 trapdoor predicates
exist.
\end{theorem}

\begin{theorem}[\cite{WC81}] Information-theoretically-secure symmetric-key message authentication codes
exist.
\end{theorem}

\begin{theorem}[\cite{NY89,Rom90}]\label{thm_SSiffOWF} Public-key signature schemes exist if and only if one-way functions
exist.
\end{theorem}

\begin{theorem}[\cite{Ren05}]\label{thm_QKD} Information-theoretically-secure
\emph{\qUKE}-protocols exist.
\end{theorem}

\noindent Because we are assuming a quantum universe, one-way
functions and trapdoor predicates\footnote{Informally, the predicate
$B(x) \in \{0,1\}$ is a(n) \emph{(unapproximable) trapdoor
predicate} if anyone can find an $x$ such that $B(x)=0$ or a $y$
such that $B(y)=1$ efficiently on a classical computer, but only one
who knows the trapdoor can, given $z$, compute $B(z)$ efficiently on
a quantum computer (this notion was introduced in Ref. \cite{GM84}).
Note that one can use a trapdoor predicate for public-key
encryption: the bit $b$ is encrypted as any $x$ such that $B(x)=b$.}
in this article (if they exist) are secure against an adversary with
a quantum computer, but are still assumed to be efficiently
computable on a classical computer; also, trapdoors are still
considered to be classical objects.\footnote{One could consider
``one-way/trapdoor quantum functions'', where the input and output
of the functions are classical or quantum, and the functions only
need to be computable efficiently on a quantum computer.
 We stick to classical one-way functions and trapdoor predicates that are quantum resistant,
candidates of which are, e.g., the trapdoor predicates underlying
some lattice-based cryptosystems (see Ref. \cite{BBD08} for more
examples).}~We also note that Theorems \ref{thm_PGiffOWF},
\ref{thm_SKEiffOWF}, \ref{thm_PKEifTP}, and \ref{thm_SSiffOWF} hold
with respect to \emph{black-box reductions}: if the theorem states
that $X$ implies $Y$, then $Y$ can be constructed from $X$, only
using $X$ as a black box, i.e., the reduction does not rely on the
specifics of how $X$ works; furthermore, the security reduction is
also a black-box one, i.e., an algorithm for breaking $X$ can be
constructed from a black box for breaking $Y$. Non-black-box
theorems of this sort are also possible (for example, see Ref.
\cite{GMW91}), but are rarely required for these kinds of results,
and indeed are not required for the theorems we quote. This is
lucky, since it guarantees us that the theorems still hold with
respect to a quantum universe.

\begin{table}[hbt]\label{tbl_CompAssumptions} \centering
\begin{tabular}{|c||c|}
\hline
Protocol class & Computational assumptions \\
\hline \hline
\OOB & none \\
\PGE & one-way functions \\
\SEB & one-way functions \\
\cUKE/\SKE  & trapdoor predicates\\
\qUKE/\MACQKE & none \\
\SIGQKE & one-way functions\\
\hline
\end{tabular}
\caption{Minimal known fundamental computational assumptions
sufficient for the existence of key establishment protocols in each
class.}
\end{table}

The theorems establish the minimal fundamental computational
assumptions known to be sufficient for the existence of protocols by
class, summarized in
Table 2.
~Public-key encryption implies one-way functions \cite{Gol04}. Thus,
the classes~\cUKE~and \SKE~ require the strongest assumption in the
table---the existence of trapdoor predicates---which reflects the
fact that it is not known how to construct any protocol in these
classes without relying on (or implying) public-key
encryption.\footnote{One might declare Table 2
 misleading, since, for example, Theorem
\ref{thm_SSiffOWF} is usually regarded merely as a plausibility
result: the construction of a
 signature scheme from an arbitrary one-way function is
relatively very inefficient. To address this issue, we note that
reasonably practical constructions are known for pseudorandom
generators, symmetric-key encryption schemes, and signature schemes
from one-way \emph{permutations} \cite{Gol01,Gol04}. Thus, even
restricting to reasonably practical schemes, the class \SKE~ still
requires the assumption of a primitive possessing a trapdoor
property, as far as we know.} To facilitate our discussion, we
summarize this point as the following conjecture:

\begin{conjecture}[Classical secret key agreement implies public-key encryption]\label{conj_RequireTrap} Every protocol
in \emph{\cUKE}~implies a trapdoor predicate (with respect to a
possibly-non-black-box reduction).
\end{conjecture}

\paragraph{\textbf{{Safest fair comparison.}}}
Most articles on quantum cryptography that appeared in the 1990s and
early 2000s stressed the fact that \MACQKE~ (respectively, \qUKE) is
the only known class of in-band \textsc{ake} (respectively,
\textsc{uke}) protocols that requires no computational assumptions.
But implicitly discarding all computational assumptions in this way
makes it impossible to have a serious discussion about the relative
merits of classical and quantum protocols for key establishment
(since any classical key-establishment protocol requires some
computational assumption). So, suppose we give classical
cryptography a fighting chance: suppose we allow only \emph{the
weakest computational assumption necessary for in-band classical key
establishment}---one-way functions.

There is good reason to do this.  Trapdoor predicates seem to be
inherently less secure than one-way functions in general.  Firstly,
trapdoor predicates easily imply one-way functions \cite{Gol04},
whereas the converse is believed not to be true.  As some evidence
for this, we note that it has been shown in Ref. \cite{IR88} that,
with respect to black box reductions (and with respect to a
classical universe), one-way functions are not sufficient (even) to
imply secret key agreement (see Remark \ref{rem_SKA}; but we have
not checked that this theorem holds with respect to a quantum
universe---in general, such classical black-box no-go theorems need
not). Secondly, using the equivalences stated in Theorem
\ref{thm_SKEiffOWF} and Theorem \ref{thm_PKEifTP}, it seems far more
likely that an efficient algorithm would be found for breaking a
public-key cryptosystem (i.e. computing a trapdoor predicate) than
breaking a symmetric-key cryptosystem (i.e. inverting a one-way
function without the trapdoor property), because the public-key
cryptosystem possesses more structure in order to embed a trapdoor
into the encryption ``function''. Quantum computers are firmly
believed not to be able to invert all one-way functions efficiently;
we state this as a conjecture:

\begin{conjecture}[One-way functions exist]\label{conj_OWFsExist} Quantum-resistant
one-way functions (computable in polynomial-time on a classical
computer) exist.
\end{conjecture}

\noindent We do not mean to suggest that quantum-resistant trapdoor
predicates do not exist (we don't know).  We do suggest, though,
that the added structure of trapdoor predicates makes it much more
likely that algorithms for the underlying problems will improve at
a more unpredictable rate
: plain one-way functions are less risky.

Even allowing one-way functions, we see that \textsc{qke} has
advantages over classical systems, beyond unconditional security.

\paragraph{{{Advantages of \textsc{{QKE}} assuming (only) one-way
functions.}}} Most of the advantages below have appeared elsewhere
in the literature in one form or another, but our presentation is
motivated differently.  The following four advantages are not
intended to be totally independent; indeed, each is just a
qualitatively different consequence of the fact that the secret key
is independent of both the initial keys and classical communication
in \textsc{qke} (and that we have taken \SKE-protocols out of the
picture).

\begin{itemize}
\item {Advantage 1:} Improved security against reveal of
initial keys
\end{itemize}

In classical cryptography, the physical nature of a cryptosystem and
protocol leads to the consideration of different types of attacks,
some more serious or more technologically difficult to mount than
others. Similarly, adversaries are often categorized by their power,
for example, \emph{passive} adversaries are considered only to be
able to read certain data that is sent along a channel, whereas
\emph{active} adversaries are assumed to have complete control over
the channel.  It is also relevant to consider precisely \emph{when}
Eve may become active; a \emph{delayed} adversary is one that
remains passive until the key establishment protocol completes, but
is active immediately afterwards.

The physical nature of a \textsc{qke} system leads to the
consideration of new kinds of attacks and adversaries.  Because of
the two different channels used, Eve can now operate differently on
these two channels.\footnote{We define ``passive'' on the quantum
channel to mean having no access, since it is difficult to formulate
a definition of ``read only'' for a quantum channel.  Measurement,
which seems necessary for reading, is an active process.}   Thus an
adversary can be defined by whether it is passive, delayed, or
active on the classical and quantum channels respectively; e.g.,
(p,p) means ``passive on both channels'' and (a,d) means ``active on
the classical channel and delayed on the quantum channel''.

With these terms in place, Table 3 shows how \qAKE-protocols have
advantages over the other classical protocols that also assume (at
most) one-way functions, for certain types of adversary; the table
indicates whether secure key can be established when the initial
keys have been revealed.  For any situation where an immediate
active attack is not deployed for whatever reason (e.g. not
technologically feasible, or not a high priority at the time), a
passive adversary who knows the initial keys loses the ability to
compromise the secret key later should she become an active attacker
later. Note that if ``\scAKE'' appeared in the leftmost column of
the table, the corresponding row of ``yes''/``no'' values would look
the same as the row corresponding to the class \qAKE.

\begin{table}[hbt] \centering
\begin{tabular}{|c||c|c|c|c|c|c|}
\hline
 & (p,p)  & (d,d) & (a,p)  & (a,d) & (a,a)\\
\hline 
\OOB & no & no  & no & no & no\\
\PGE & no & no & no & no & no\\
\SEB  & no & no &no & no & no \\
\qAKE & yes & yes & yes & yes & no \\
\hline
\end{tabular}
\caption{Security against reveal of initial keys. The entries
(yes/no) of the chart indicate whether the secret key generated from
the key establishment protocol is secure under the reveal of either
Alice's or Bob's initial key for the given adversary (see the main
text for an explanation of the notation used to define the
adversaries).  The class \SKE~ does not appear, since we are not
assuming trapdoor predicates (and there is no known \SKE-scheme that
does not imply trapdoor predicates).}
\end{table}

Note that, in order to break a \qAKE-protocol---or, more precisely,
break the cryptosystem that comprises the \qAKE-protocol---Eve,
knowing all the initial keys, can mount an active and sustained
``man-in-the-middle'' attack; furthermore, for a \MACQKE-system, the
active attack must occur during the first instance of the protocol
(as any subsequent instance will use different and independent
initial keys).  In large networks, this may pose a considerable
challenge for Eve, depending on when she learns the initial keys and
whether the connections among users are fixed or ad-hoc.

\begin{remark}[Perfect forward secrecy]  Note that Advantage 1 is
different from perfect forward secrecy, a \emph{much} weaker notion
referring to whether secret keys established in \emph{past sessions}
(with old initial keys no longer stored on Alice and Bob) are secure
once \emph{current} initial keys are revealed. While
\emph{\qAKE}-protocols certainly have perfect forward secrecy,
Bernstein \cite{Ber09} has noted that well-implemented
\emph{\PGE}-protocols do, too.
\end{remark}

\begin{itemize}
\item {Advantage 2:} Reduced dependence on out-of-band actions
\end{itemize}

Because a \MACQKE-protocol generates secret key that is independent
of the initial keys and the classical communication, initial keys
can be smaller in the \MACQKE-protocol than in an \OOB-protocol,
i.e., less initial entropy is needed to prime the system.  Also, a
\MACQKE-system may require fewer subsequent out-of-band actions for
refreshing initial keys, compared to \PGE- and \SEB-systems (at the
very least because the latter are more vulnerable to
initial-key-reveal attacks---see above).

\begin{itemize}
\item {Advantage 3:} Reduced dependence on trusted third parties
\end{itemize}

In a network, key establishment can be done in a mediated fashion,
via a trusted \emph{key distribution centre}, whose job is to give
\emph{session keys} to Alice and Bob so that they may communicate
securely.  As part of the setup, every user in the network,
including Alice and Bob, shares an initial key (established out of
band) with the key distribution centre; in principle, these initial
keys may be asymmetric or symmetric.  An example of such a system is
Kerberos, where the initial keys are symmetric, and, upon request by
either Alice or Bob, the key distribution centre generates a
symmetric key and sends it (encrypted using the initial keys) to
Alice and Bob, who then use it to encrypt and decrypt messages
between each other.

Quantum key establishment may also be done in a mediated fashion, so
that the channels connecting Alice to Bob go through a key
distribution centre, which gives Alice and Bob a session key to be
used as a symmetric initial key in a \MACQKE-protocol.

If trapdoor predicates are not assumed to exist, then any classical
mediated key establishment system must use symmetric initial keys;
this is because the key distribution centre must send keys to Alice
and Bob, and these keys must be, at least partially, encrypted
(assuming the key distribution centre is not to play an active part
in the communication between Alice and Bob). Similarly, the session
keys must be symmetric keys, too.

Comparing any classical mediated key establishment system to one
where Alice and Bob use their symmetric session keys as initial keys
in a \MACQKE-protocol, we see that, in the quantum case, Alice and
Bob do not need to trust the key distribution centre after their key
establishment protocol is complete.  By contrast, in the classical
case, the key distribution centre must always be trusted, since it
knows the keys that Alice and Bob use to communicate securely.  As
well, Alice and Bob may be able to decouple themselves completely
from the key distribution centre after their first \MACQKE-session.
Thus, any compromise of the key distribution centre after the first
\MACQKE-session does not necessarily affect Alice and Bob.

\begin{itemize}
\item {Advantage 4:} Long-term security from short-term security
\end{itemize}

The secret key generated by any \qAKE-protocol will be
information-theoretically secure even if the authentication
algorithm is broken in the short term---as long as the break occurs
after the key establishment protocol is completed. We may refer to
this as ``conditional information-theoretic security''.  This allows
for the use of authentication algorithms that are perhaps less
secure in the long term but are easier to manage with regard to
initial keys, i.e., public-key algorithms. Note that any
\SIGQKE-system has the extra advantage over a \MACQKE-system that it
is less susceptible to running out of authentication key due to
noise or eavesdropping, because there is no practical limit on how
many classical messages may be authenticated. In other words, using
public-key authentication guards against at least one type of
denial-of-service attack.

Also, Alice and Bob may not need to rely on the same type of
authentication used for the first \qAKE-session for subsequent
\qAKE-sessions, i.e., for the first session, Alice and Bob may
execute a \SIGQKE-protocol, but, for all subsequent sessions (in
principle, i.e., in the absence of sufficiently heavy adversarial
action or noise), they may execute a \MACQKE-protocol.  Two
potential advantages of such a two-phase system are that (1)
subsequent key establishment sessions may run faster (since the
symmetric-key algorithms may be more efficient than public-key
algorithms for the required level of security) and (2) subsequent
key establishment sessions may not need to rely on any computational
assumptions.

If quantum computers can be assumed not to exist in the short term,
i.e., for the service-lifetime of the public keys, then one can even
use public-key signature schemes whose security relies on the
assumption of hardness of factoring and the discrete logarithm
problem for classical computers.

We believe that its ability to derive long-term from short-term
security, also known as \emph{everlasting security},\footnote{The
term ``everlasting security'' has been used in the context of the
bounded storage model (see, e.g., Ref. \cite{CM97}), where, e.g., it
describes the case where encryption is secure even if the adversary,
at some later time, learns the pre-shared symmetric key, as long as,
at the time of transmission of the ciphertext, the adversary has
bounded storage capability (see Ref. \cite{DR02}).  The term seems
equally well suited to \textsc{qke}.} may be the most attractive
aspect of \textsc{qke} systems from a security perspective.

\paragraph{\textbf{{The baby...}}}
The advent of public-key cryptography revolutionized secure
telecommunications, by vastly simplifying the problems of key
distribution and key management:  Alice and Bob no longer needed to
pre-share a symmetric key. Instead, Alice could publish her own
public key, and that would be sufficient for her to receive
encrypted messages from anyone who got a hold of it.

Of course, ``publishing'' a public key is easier said than done, but
public-key cryptography helps solve this problem, too.  A signature
scheme can be used in conjunction with a network of trusted third
parties to help Bob be certain that he has Alice's legitimate public
key.\footnote{On the Internet, this works as follows.  Bob's
web-browser comes from the manufacturer pre-loaded with the public
key of a trusted third party Charlie. When Bob wants to communicate
with Alice, she shows Bob a certificate which contains her purported
public key and Charlie's signature of the certificate, which also
contains Alice's name (and other uniquely identifying and
publicly-agreed-upon details about Alice). Bob checks that Alice's
public key is valid by verifying Charlie's signature using the
pre-loaded public key. In this context, signature schemes are said
to offer ``manageable persistence'' (via digital signature) of the
binding of a name and a key \cite{AL03}.}\label{foot_PKI}  This is
probably the reason Rivest \cite{Riv90} wrote, ``The notion of a
digital signature may prove to be one of the most fundamental and
useful inventions of modern cryptography.''

\paragraph{\textbf{{...the bathwater.}}}
There is a price to pay for the advantages of a public-key
infrastructure.  Security necessarily depends on assumptions about
the hardness of certain mathematical problems; proofs that such
problems are actually hard seem to be beyond the reach of
theoretical computer scientists.

After Peter Shor discovered an efficient quantum algorithm for
factoring and computing discrete logarithms in 1994, \textsc{qke}
protocols, the earliest of which dates back to 1984, received
renewed interest. Most literature on \textsc{qke} that appeared in
the 1990s and early 2000s focussed on protocols in the class
\MACQKE.  And rightfully so:  it is remarkable that symmetric
initial keys can be expanded into much larger, independent, and
information-theoretically secure secret keys in band by exploiting
quantum mechanics.   As such, these articles, through their
reference to Shor's discovery, may have been seen as suggesting that
all computational assumptions should be jettisoned at the earliest
opportunity---for who knew what problems might next succumb to the
power of a quantum computer?

\paragraph{\textbf{{A new spin on quantum cryptography.}}}
It was known (though perhaps not widely) that insisting on
unconditional security was not the only way forward in order to
ensure reasonable security against quantum attacks.  It was evident
that public-key signature schemes could be used to authenticate the
classical channel in a \textsc{qke} protocol, and that such a system
would have some attractive features; this idea first appeared in the
literature in Ref. \cite{PPS07}.
Indeed, in light of Theorem \ref{thm_SSiffOWF} and Table 2
, and assuming 
Conjecture \ref{conj_OWFsExist} is true, this idea becomes rather
more striking:

\begin{itemize}
\item \emph{Quantum cryptography is the only known way to achieve (quantum-resistant) private
communication in a public-key infrastructure with the minimal
computational assumptions}.
\end{itemize}

\noindent (If in addition Conjecture \ref{conj_RequireTrap} is true,
then the word ``known'' can be dropped.)  In other words, with some
abuse of the metaphor, quantum cryptography potentially allows us to
throw out some of the bathwater---i.e., primitives with a trapdoor
property---while keeping most of the baby---i.e., authenticated
encryption without symmetric initial keys---and no classical scheme
is known to accomplish this. At the very least, quantum cryptography
certainly allows us to sidestep the question of the necessity of
trapdoor predicates for secret key agreement (or trapdoor functions
for trapdoor predicates \cite{GMR01}). We view this as strengthening
the case for signed \textsc{qke}.

\paragraph{\textbf{{If public-key encryption exists...}}}
If trapdoor predicates do exist and are secure in the long term, we
note that Advantages 1 through 4 can variously be achieved by
\SKE-protocols to at least some degree. However, in this case,
\textsc{qke} protocols may have other advantages over classical
ones.  Because the secret key $s$ generated in a \qAKE-protocol is
independent of the classical communication $c$, there is no
mathematical way to connect these two quantities
or---{attribute}---the secret key to Alice's and Bob's publicly
readable discussion; we say that the secret key is
\emph{nonattributable}.\footnote{In Ref. \cite{Bea02}, Beaver
discusses ``deniability'' (see Refs \cite{CDNO96,KPK08}) of
\textsc{qke}, which is similar to nonattributability.  However, in
that paper, it is assumed that Alice and Bob keep a record of their
qubit-measurement outcomes (often called ``raw key bits'') made
during the protocol and that, if Alice and Bob are to deny that a
particular secret key was established, this record must be
consistent with any measurements made by an eavesdropper, i.e.,
someone who is forcing Alice or Bob to reveal the secret key (or the
plaintext encrypted by it). We assume that Alice and Bob do not keep
such records and that it is sufficient that the forcer cannot
provide evidence that attributes a particular secret key to the
classical communication; any measurement on the quantum channel that
the forcer made is not publicly verifiable, so we do not view its
outcome as part of the public record.  In other words, in our model,
Alice and Bob need not provide evidence to support their (tacit)
denial.  Incidentally, Beaver concludes that the standard
\textsc{qke} protocols do not provide deniability in his model.}

There are two ways in which a secret key may be considered
attributable: it is attributable to Alice's and Bob's public
discussion (through its dependence on the classical communication)
and it is attributable to Alice and/or Bob (because they
participated in the classical communication). For the former way, we
just use the term \emph{attributable} to describe the secret key;
for the latter way, we say the secret key is
\emph{party-attributable}. If the classical communication is
authenticated via a signature scheme, then the secret key may be
party-attributable in a provable way, or \emph{provably
party-attributable}.  If the secret key is subsequently used in an
encryption scheme to encrypt a plaintext, then we say that the
plaintext is (party- or provably party-) attributable whenever the
secret key is.

Because \qAKE-protocols do not produce an attributable secret key, a
\SIGQKE-protocol may be used in composition with a one-time pad
encryption scheme, and then the secret key (and hence the plaintext)
would never be attributable.  No totally classical scheme can
achieve the same thing, i.e., non-party-attributable, public-key,
secure communication.

For symmetric-key ciphers where the bit-length of the secret key is
much smaller than the bit-length the message (e.g., \textsc{aes}),
the cipher itself provides a subroutine for recognizing the secret
key (i.e., if a candidate secret key $s'$ decrypts the ciphertext to
something sensible, then with high probability $s'$ equals the
actual secret key). If the secret key was produced by a
\SIGSKE-protocol, then the secret key (and hence the plaintext) are
provably party-attributable given the secret key; however, if the
secret key was produced by a
 \SIGQKE-protocol, it is not attributable at all.  This
 is a potential advantage of using \textsc{qke} to generate
 \textsc{aes} keys.

\paragraph{\textbf{{Closing Remarks.}}}
Recall the objections to \textsc{qke} that we listed earlier (see
Page \pageref{objections}).  We have addressed Objection 4 early on,
by highlighting the fundamental distinction between in-band and
out-of-band key establishment protocols.  We believe there exist (or
will exist) applications where in-band generation of entropy is
desirable.

Objections 2 and 3 both propose using (potentially very long)
symmetric initial keys in \OOB~or \PGE~protocols. We have presented
a considerable list of advantages that \textsc{qke} has over these
protocols.

Objection 1 is the strongest one, but it relies on the computational
assumption of a trapdoor predicate, which (until any lower bounds
are proven) incurs risk when public-key encryption is used for
long-term secrets.  The field of quantum algorithms is still
relatively young, so it is probably unwise to assume any particular
candidate trapdoor predicate with a particular set of parameters is
secure (the recent discovery of a subexponential-time quantum
algorithm for elliptic curve isogenies supports this perspective
\cite{CJS10}). However, in addition to these standard
counter-arguments for Objection 1, we have shown that \textsc{qke}
may offer the benefit of nonattributability in scenarios where no
purely classical scheme can. We also note that it is conceivable
that, in the future, a \qAKE-system may be more efficient (i.e. have
a higher secret key rate) than a \SKE-system, as public-key
encryption is known to be rather slow. As well, \qAKE-systems may be
more cost-effectively resistant to side-channel attacks, which are
notoriously difficult to defend against in the classical world.

The debate on the merits of \textsc{qke} may have suffered from a
focus on unconditional security, which may have given the impression
that it is of no value to practical cryptography.  The message from
classical cryptographers has been loud and clear: the pre-sharing of
symmetric keys is costly and thus to be avoided in the majority of
key-establishment applications: e.g., Paterson et al. \cite{PPS07}
wrote, ``[Quantum key establishment], when unconditionally secure,
does not solve the problem of key distribution. Rather, it
exacerbates it, by making the pre-establishment of symmetric keys a
requirement.'' They also wrote, ``It is likely that using
[\textsc{qke}] with public key authentication [...] has security
benefits [...]. However, [\textsc{qke}] loses much of its appeal in
[this setting], as the overall system security is no longer
guaranteed by the laws of quantum physics alone.''  Our article is
completely in accordance with the former comment and, with regard to
the latter comment, expands on the ``benefits'' of signed
\textsc{qke} in order to bolster its ``appeal''. As such, we hope to
have firmed up the middle ground between unconditionally-secure
\textsc{qke} and computationally-secure classical key establishment
in the ``quantum debate''.

\bibliographystyle{unsrt}


\end{document}